\documentclass[twocolumn,showpacs,preprintnumbers,pre,aps,10pt]{revtex4-1}
\usepackage{bm}
\usepackage{amssymb,amsmath}
\usepackage{graphicx}
\usepackage{natbib}
\usepackage{program}
\usepackage{color}
\definecolor{darkgreen}{rgb}{0.15,0.5,0.15}
\definecolor{darkblue}{rgb}{0.15,0.15,0.5}

\usepackage[backref,breaklinks,colorlinks,citecolor=blue]{hyperref}

\newcommand{\pluseq}{\mathrel{+}=}

\newcommand{\Ptensor}{\mathsf{P}}  

\begin{document}
\title[NLSE SPH]{Numerical solution of the non-linear Schr\"odinger equation using smoothed-particle hydrodynamics}
\author{Philip Mocz}\email{pmocz@cfa.harvard.edu}
\affiliation{Harvard-Smithsonian Center for Astrophysics, 60 Garden Street, Cambridge, MA 02138, USA}
\author{Sauro Succi}\email{succi@iac.cnr.it}
\affiliation{Istituto per le Applicazioni del Calcolo, CNR, Viale del Policlinico 137, I-00161, Roma, Italy \\
Institute of Applied Computational Science, Harvard School of Engineering and Applied Sciences, Northwest B162, 52 Oxford Street, Cambridge, MA 02138, USA}

\date{\today}
\begin{abstract}
We formulate a smoothed-particle hydrodynamics numerical method, traditionally used for the Euler equations for fluid dynamics in the context of astrophysical simulations, to solve the non-linear Schr\"odinger equation in the Madelung formulation. The probability density of the wavefunction is discretized into moving particles, whose properties are smoothed by a kernel function. The traditional fluid pressure is replaced by a quantum pressure tensor, for which a novel, robust discretization is found. We demonstrate our numerical method on a variety of numerical test problems involving the simple harmonic oscillator, soliton-soliton collision, Bose-Einstein condensates, collapsing singularities, and dark matter halos governed by the Gross-Pitaevskii-Poisson equation. Our method is conservative, applicable to unbounded domains, and is automatically adaptive in its resolution, making it well suited to study problems with collapsing solutions.
\end{abstract}
\smallskip
\maketitle

\section{Introduction}\label{sec:intro}

Quantum mechanics is one of the basic pillars of modern physics. The Schr\"odinger equation describes the quantum mechanical evolution of the wavefunction of a particle over time. The non-linear Schr\"odinger equation (NLSE), also called the Gross-Pitaevskii equation, is a non-linear extension of the Schr\"odinger equation, which describes the ground state of a quantum system of identical bosons using a single-particle wavefunction approximation and a pseudopotential model for interaction. It is ideal for describing a Bose-Einstein condensate (BEC): dilute gas of bosons in a low-temperature state very close to absolute zero. BECs were first predicted in the early days of quantum theory by Bose and Einstein in 1924-1925. The first realization in the laboratory was achieved in 1995 \citep{bradley1995evidence,anderson1995observation}, which marked a new era in atomic, molecular and optical (AMO) physics and quantum optics \citep{bao2014mathematical}. The NLSE has applications and extensions to entirely different physical systems as well, including the propagation of light in non-linear fiber optics \citep{anderson1983variational}, Langmuir waves in plasmas \citep{gupta1981coupled}, and self-gravitating BEC models for dark matter, governed by the Gross-Pitaevskii-Poisson equations \citep{chavanis2011mass}.

The NLSE is challenging to solve and almost always requires numerical solutions. Ongoing research has led to the development of a variety of methods to solve these systems in various contexts, such as those for solving time-evolution of BEC systems \citep{bao2003numerical,bao2013mathematical,antoine2014gpelab,bao2014mathematical,bao2014mathematical,santos2014comparison}
and obtaining their ground states \citep{Dalfovo96thecondensate,bao2003ground,bao2004computing,bao2014ground}. These methods solve for the solution to the NSLE in the standard form, and typically employ finite-difference, finite-element, or spectral methods. Other non-standard methods for solving quantum systems have been proposed as well, such as lattice Boltzmann \citep{succi1996numerical,palpacelli2008quantum} and unitary qubit lattice algorithms \citep{yepez1998lattice,yepez2001quantum}. Each method has different strengths and limitations when applied to different systems \citep{minguzzi2004numerical}.

We propose a novel, conservative numerical approach for solving the NLSE that is quite different from the standard approaches. We solve the NLSE in Madelung hydrodynamic form, using a smoothed particle hydrodynamics (SPH) algorithm. The Schr\"odinger equation, as well as the NLSE, can be reformulated under the Madelung transformation to take a different form that resembles the fluid equations \citep{tsekov2009dissipative}. The equation in Madelung form describes the evolution of the quantum probability density of the wavefunction under a quantum ``pressure'' tensor, and is equivalent to the standard form.

SPH is a particle-based method for computational fluid dynamics. It was originally invented to simulate polytropic stellar models under non-axisymmetric conditions \citep{gingold1977smoothed,lucy1977numerical}. It has since been extended and coupled with additional physical processes and plays a central role in astrophysical and cosmological simulations \citep{monaghan1992smoothed,springel2010smoothed,price2012smoothed}. SPH operates independently of any grid, unlike finite-difference, finite-volume, or finite-element methods, and interactions between volume elements, such as the pressure gradient, are represented as a force between particles. The method is purely Lagrangian, meaning that interactions and derivatives are evaluated in a coordinate system attached to a moving fluid element. The two fundamental ideas of SPH are (1) to evolve the positions and velocities of particles according to the calculation of the forces on each particle at each time step, and (2) to use an interpolating/smoothing kernel to calculate forces and spatial derivatives.

SPH has some desirable inherent features for quantum systems. The method is conservative, so the normalization condition on the wave function is preserved to machine precision. Also, the SPH method also has no domain restrictions; the wave function is free to travel anywhere in physical space, which is an advantage grid-based methods do not possess. The Lagrangian nature of SPH also makes the method useful for the study of highly dynamic solutions, such as rapidly rotating wavefunctions. Furthermore, the SPH method is automatically adaptive in its resolution, making it possible to easily resolve collapsing features in the solution, and this is one of the primary reasons it often is used in cosmological simulations of structure formation.

Our method may be extended in a relatively straightforward manner to various other quantum systems, such as multi-component, rotational, dipolar, or spin-orbit coupled BECs; to Euler-Korteweg systems for capillary fluids; or to the study of the semi-classical limit of the Schr\"odinger equation \citep{berry1972semiclassical,fibich2005new}. Such applications are left to future study.

The remainder of this paper is organized as follows. In Section~\ref{sec:theory} we discuss the theory of the NLSE and the Madelung formulation. In Section~\ref{sec:method} we describe our numerical SPH method for the NLSE. In Section~\ref{sec:results} we demonstrate the accuracy and results of our method on some numerical test problems. In Section~\ref{sec:conc} we offer concluding remarks and list physics areas of applications for the numerical method.

\section{Theory}\label{sec:theory}

The Schr\"odinger equation for quantum mechanics may be written in dimensionless form ($\hbar=1$) as
\begin{equation}
i\partial_t \psi(\mathbf{x},t) = \left[ -\frac{1}{2}\nabla^2 + V(\mathbf{x}) \right]\psi,\,\,\,\,\,\, \mathbf{x}\in\mathbb{R}^3,\,\,\,\,\,\,t>0 .
\label{eqn:SE}
\end{equation}

The dynamics of a BEC is well described by the NLSE. The NLSE in dimensionless form \citep{bao2013mathematical} may be written as 
\begin{equation}
i\partial_t \psi = \left[ -\frac{1}{2}\nabla^2 + V + g \lvert\psi\rvert^2 \right]\psi .
\label{eqn:NLSE}
\end{equation}
In this form, the normalization is $\int \lvert\psi\rvert^2 \,d^3\mathbf{x}= 1$. $g\in\mathbb{R}$ is treated as an arbitrary dimensionless parameter which measures the strength of nonlinear interactions.

Under the Madelung transformation \citep{tsekov2009dissipative}, the NLSE resembles the fluid equations:
\begin{equation}
\partial_t \rho + \nabla \cdot (\rho \mathbf{u}) = 0 ,
\label{eqn:madelung1}
\end{equation}
\begin{equation}
\partial_t \mathbf{u} + \mathbf{u} \cdot \nabla \mathbf{u} = -\frac{1}{\rho} \nabla \cdot\Ptensor - \frac{g}{\rho} \nabla \frac{\rho^2}{2} - \nabla V ,
\label{eqn:madelung2}
\end{equation}
where $\rho = \lvert\psi\rvert^2$ is the quantum probability density of the wavefunction, and $\mathbf{u}=\nabla\theta$, where $\psi \equiv \lvert\psi\rvert {\rm exp}\left(i\theta(\mathbf{x},t)\right)$. The variable
\begin{equation}
\Ptensor = -\frac{1}{4}\rho \nabla \otimes \nabla \ln \rho
\label{eqn:Ptensor}
\end{equation}
is the quantum pressure tensor. Often in the literate, the quantum pressure term is instead written in terms of a quantum potential $Q=-\frac{1}{2}\frac{\nabla^2\sqrt{\rho}}{\sqrt{\rho}}$, but it turns out that for the purposes of discretizing the equations to obtain an SPH method, the pressure tensor formulation is more useful.

Optionally, one may add artificial damping to the equations, with damping parameter $\gamma$, as
\begin{equation}
\partial_t \mathbf{u} + \mathbf{u} \cdot \nabla \mathbf{u} = -\frac{1}{\rho} \nabla \Ptensor - \frac{1}{\rho} \nabla \frac{g \rho^2}{2} - \nabla V - \gamma\mathbf{u} .
\label{eqn:madelung2damp}
\end{equation}
This can be useful to bring solutions to a steady state. Such a term describes dissipative quantum systems \citep{tsekov2009dissipative}, where the system loses energy with time. For our purposes, the damping term is useful in relaxing an arbitrary wave function to it's ground state.

The NLSE has been coupled with self-gravity to form the Gross-Pitaevskii-Poisson equation. These equations describe models for BEC dark matter halos \citep{chavanis2011mass}. In this case, the potential is computed from the wavefunction according to Poisson's equation:
\begin{equation}
\nabla^2 V = M^24\pi G\rho.
\label{eqn:poisson}
\end{equation}
(Note, there is a factor of $M^2$, where $M$ is the total mass of the system, because we are using units where $\rho = \lvert\psi\rvert^2$ is dimensionless.)

Another variant of the NLSE is the focusing NLSE
\begin{equation}
i\partial_t \psi(\mathbf{x},t) = \left[ -\nabla^2 - \lvert \psi \rvert^{2\sigma} \right]\psi,\,\,\,\,\,\, \mathbf{x}\in\mathbb{R}^d
\label{eqn:sf}
\end{equation}
where solutions exist that self focus and become singular in finite time for the case $\sigma d \geq 2$. Numerically solving such a blowup solution is challenging because the spatial and temporal gradients grow arbitrarily large while small perturbations may arrest the critical collapse. Standard grid methods break down and more sophisticated methods, which involve dynamical rescaling, have been resorted to in order to handle blowup solutions \citep{fibich2003discretization}.

\section{Numerical Method}\label{sec:method}

We discretize the quantum probability density $\rho = \lvert\psi\rvert^2$ of the wavefunction as a collection of $N$ particles. Each particle is assigned a ``mass'' $m_j=1/N$ so that we satisfy the normalization condition $\int \lvert\psi\rvert^2 \,d^3\mathbf{x}\simeq \sum_j m_j=1$ to machine precision.

We wish to solve for the dynamics of $\rho$ using the Madelung formulation. Equation~\ref{eqn:madelung1} is just a statement about the conservation of the normalization condition, which is automatically satisfied by our discretization. Equation~\ref{eqn:madelung2} describes the equation of motion of the particles. The left-hand side is a convective (Lagrangian) derivative of the velocity: $\frac{d\mathbf{u}}{dt}$, namely, the acceleration.

The main goal of the SPH method is to evaluate the acceleration of each particle and update the velocities and positions of the particles with each time step using an integrator scheme (Section~\ref{sec:leap}).

With SPH, the value of a field at any point in the domain is obtained by smoothing out the values associated with the particles. Consider the (trivial) identity:
\begin{equation}
A(\mathbf{x}) = \int A(\mathbf{x}^\prime) \delta(\mathbf{x}-\mathbf{x}^\prime) \, d\mathbf{x}^\prime
\label{eqn:A}
\end{equation}
where $A(\mathbf{x}):\mathbb{R}^3\mapsto \mathbb{R}$ is any arbitrary function and $\delta(\mathbf{x})$ is the Dirac delta function. In the SPH scheme, $\delta$ is replaced with an approximation: a smoothing kernel $W(\mathbf{x}; h)$, where $h$ is the smoothing length scale. The smoothing kernel must have the properties
\begin{equation}
\int W(\mathbf{x}; h)\,d^3\mathbf{x} = 1,
\end{equation}
\begin{equation}
\displaystyle\lim_{h\to 0} W(\mathbf{x};h) \to \delta(\mathbf{x}).
\end{equation}
and must be non-negative and invariant under parity. 

For our purposes, we choose the Gaussian kernel:
\begin{equation}
W(\mathbf{x};h) = \left(\frac{1}{h\sqrt{\pi}}\right)^3 {\rm exp} \left(-\|\mathbf{x}\|^2/h^2\right)
\end{equation}
where $h$ is a smoothing-length parameter. Alternate choices include cubic-splines, which have compact support and can make pairwise-interaction calculations more efficient. For our purposes, we will most often use a fixed value for $h$, but more sophisticated formulations of SPH exist which use adaptive values based on particle number density \citep{nelson1994variable,springel2010smoothed} (see Section~\ref{sec:adaptive}).

Hence, an approximation to the field $A(\mathbf{x})$ in Equation~\ref{eqn:A} is
\begin{equation}
A(\mathbf{x}) \simeq \int A(\mathbf{x}^\prime) W(\mathbf{x}-\mathbf{x}^\prime) \, d\mathbf{x}^\prime
\end{equation}
To this equation, we apply a second approximation, namely discretization: we sum over the $N$ particles. The SPH approximation to $A(\mathbf{x})$ is thus
\begin{equation}
A(\mathbf{x}) \simeq \sum_j \frac{m_j}{\rho_j} A(\mathbf{x}_j)W(\mathbf{x}-\mathbf{x}_j;h) 
\label{eqn:A2}
\end{equation}
where $\mathbf{x}_j$ is the location of particle $j$ in physical space and $\rho_j=\rho(\mathbf{x}_j)$ (calculated via Equation~\ref{eqn:rho}).

Similarly, gradients of fields can be approximated as follows
\begin{equation}
\nabla A(\mathbf{x}) \simeq \sum_j \frac{m_j}{\rho_j} A(\mathbf{x}_j)\nabla W(\mathbf{x}-\mathbf{x}_j;h) .
\label{eqn:Aderiv}
\end{equation}
Note that the gradient operator shifts to the kernel, whose derivative is analytically known.

\subsection{Calculating density}\label{sec:rho}

Given a distribution of $N$ particles in physical space, the density $\rho_i$ at each particle $i$ is required to estimate any field quantity, because it appears in Equation~\ref{eqn:A2}. This is calculated straightforwardly by substituting $A(\mathbf{x})=\rho(\mathbf{x})$ into Equation~\ref{eqn:A2}:
\begin{equation}
\rho_i =  \sum_j m_j W_{ij}
\label{eqn:rho}
\end{equation}
where we have defined
\begin{equation}
W_{ij} \equiv W(\mathbf{x}_i-\mathbf{x}_j;h) .
\end{equation}

\subsection{Calculating pressure tensor}\label{sec:press}

In classical fluid dynamics applications, the pressure $P_i$ at a particle location is calculated using an equation of state, which depends on quantities such as the density and/or the internal energy of the fluid. One example is the polytropic equation of state $P = k\rho^{1+1/n}$, where $k$ is a constant and $n$ is the polytropic index.

In the case of quantum mechanics, the pressure is replaced by a symmetric pressure tensor (Equation~\ref{eqn:Ptensor}), which is a non-local quantity because it depends on gradients of the density.

First, the derivatives of the density field are calculated at the location of each particle, along the lines of Equation~\ref{eqn:Aderiv}:
\begin{equation}
\partial_x \rho_i = \sum_j m_j \partial_x W_{ij} .
\end{equation}
Similarly for $\partial_y$, $\partial_z$.

The second derivatives can be calculated in similar fashion as
\begin{equation}
\partial_{xy} \rho_i = \sum_j m_j \partial_{xy} W_{ij} ,
\end{equation}
and similarly for $\partial_{xx}$, $\partial_{xz}$, $\partial_{yy}$, $\partial_{yz}$, $\partial_{zz}$.
But this is not the only discretization. They may also be calculated using the (more accurate) second-order discretization \citep{chaniotis2002remeshed}:
\begin{equation}
\partial_{xy} \rho_i = \sum_j \frac{m_j}{\rho_j}\left(\rho_j-\rho_i\right) \partial_{xy} W_{ij} ,
\end{equation}
Other alternate discretizations exist in the literature as well, such as the difference scheme, and are widely used for applications such as heat conduction \citep{cleary1998modelling,fatehi2008discretization}.

Finally, the components of the quantum pressure tensor of Equation~\ref{eqn:Ptensor} are computed as:
\begin{equation}
\Ptensor_{i,xy} = \sum_j \frac{m_j}{\rho_j} \frac{1}{4}\left[ \frac{(\partial_x \rho_j)(\partial_y \rho_j)}{\rho_j} - \partial_{xy} \rho_j \right] W_{ij}
\end{equation}
and similarly for the rest of the components of the tensor. This is on of the main equations of our paper, which provides a robust discretization for the quantum pressure tensor that yielded well-behaved solutions in all our test problems. 

We note here an analogy between the quantum pressure tensor $\Ptensor = -\frac{1}{4}\rho \nabla \otimes \nabla \ln \rho$ and the equations for inviscid capillary fluids. Such classical, non-ideal fluids can be described by the Euler-Korteweg equations, which have the form:
\begin{equation}
\frac{d\mathbf{u}}{dt} = \nabla\left(
\kappa(\rho)\Delta^2 \rho + \frac{1}{2}\kappa^\prime(\rho) \lvert \nabla \rho\rvert^2
\right)
\end{equation}
Taking $\kappa(\rho) = 1/(4\rho)$ leads back to the NLSE. In the classical case, the pressure tensor would act as a surface tension term, and its sign would act to add decoherence into the solution (which is a typical quantum phenomenon). An opposite sign would lead to cohesion.

\subsection{Calculating acceleration}\label{sec:acc}

A key guiding principal in formulating an SPH method for obtaining robust results is to choose discretizations for the forces the particles experience such that the forces between pairwise particles obey Newton's third law, i.e., are equal and opposite. This allows for the particles to quasi-regularize as they sample the true solution of the field \citep{price2012smoothed}. This leads to better-than-random-Monte-Carlo sampling in the reconstruction of field quantities.

In the standard SPH formulation with scalar fluid pressure, the acceleration of a particle due to the pressure gradient $-\frac{1}{\rho}\nabla P$ is calculated in symmetric fashion as:
\begin{equation}
\frac{d\mathbf{u}_i}{dt} = -\sum_j m_j \left( \frac{P_i}{\rho_i^2} + \frac{P_j}{\rho_j^2} \right) \nabla W_{ij} .
\label{eqn:accsph}
\end{equation}

The force due to the pressure tensor can also be calculated in a similar fashion:
\footnotesize
\begin{align}
& \frac{d\mathbf{u}_i}{dt} = -\sum_j m_j \Big[  \nonumber\\
&\,  \left( \frac{\left[ \Ptensor_{i,xx},\, \Ptensor_{i,xy},\, \Ptensor_{i,xz} \right]}{\rho_i^2} + \frac{\left[ \Ptensor_{j,xx},\, \Ptensor_{j,xy},\, \Ptensor_{j,xz} \right]}{\rho_j^2} \right) \cdot \nabla W_{ij}  , \nonumber\\
&\,  \left( \frac{\left[ \Ptensor_{i,yx},\, \Ptensor_{i,yy},\, \Ptensor_{i,yz} \right]}{\rho_i^2} + \frac{\left[ \Ptensor_{j,yx},\, \Ptensor_{j,yy},\, \Ptensor_{j,yz} \right]}{\rho_j^2} \right) \cdot \nabla W_{ij}  , \nonumber\\
&\,  \left( \frac{\left[ \Ptensor_{i,zx},\, \Ptensor_{i,zy},\, \Ptensor_{i,zz} \right]}{\rho_i^2} + \frac{\left[ \Ptensor_{j,zx},\, \Ptensor_{j,zy},\, \Ptensor_{j,zz} \right]}{\rho_j^2} \right) \cdot \nabla W_{ij}  , \nonumber\\
&  \Big] .
\label{eqn:main}
\end{align}
\normalsize
Obeying Newton's third law between pairwise particles is one of the key reasons we chose to discretize a version of the Madelung equations that uses the pressure tensor. Note the inner square brackets here denote a vector, and we have written out the entries explicitly.

We can also calculate the additional acceleration due to the non-linear term $-\frac{1}{\rho}\nabla \frac{g \rho^2}{2}$ in Equation~\ref{eqn:madelung2} by setting $P_i = \frac{g \rho_i^2}{2}$ in right-hand side expression of Equation~\ref{eqn:accsph}. 

Optionally, one may artificially damp the solutions by adding the damping term $\gamma \mathbf{u}_i$ to the right hand side of Equation~\ref{eqn:main}. Damping is useful for obtaining steady-state solutions: eigenstates and/or ground-states of the system. Is is also useful for generating initial conditions.

\subsection{Leap frog time integration}\label{sec:leap}

The particles are initialized with positions and velocities dictated by the initial conditions of the problem (in problems that use damping to reach a steady state, the initial conditions can be chosen to be a random or uniform distribution of particles with $0$ velocity; the particle configurations of such steady state solutions can then be used for problems with dynamics). Then, equation~(\ref{eqn:madelung2}) may be solved with a time integration method, such as Runge-Kutta or leap frog. The leap frog method is often preferred because it is explicit and symplectic. In our implementation, we use the second-order leap frog scheme as follows:
\begin{equation}
\mathbf{u}(t+\Delta t/2) = \mathbf{u}(t-\Delta t/2) + \mathbf{a}(t)\Delta t
\label{LF1}
\end{equation}
\begin{equation}
\mathbf{x}(t+\Delta t) = \mathbf{x}(t) + \mathbf{u}(t+\Delta t/2)\Delta t
\end{equation}
that is, we calculate positions and velocities at interleaved time points. We note that at the start of the simulation we only know the initial conditions $\mathbf{x}(0)$ and $\mathbf{u}(0)$, and must use first-order Euler to step back half a time step and find $\mathbf{u}(-\Delta t/2)$.

To find the velocities at the same time step intervals as the positions, we use the approximation:
\begin{equation}
\mathbf{u}(t) = \frac{1}{2}\left(\mathbf{u}(t-\Delta t/2)+\mathbf{u}(t+\Delta t/2)\right) .
\end{equation}

\subsection{Pseudocode}\label{sec:pseudo}

A pseudocode of the main loop of the SPH algorithm is shown below.

\begin{program}{\color{darkblue}{|Main Loop|}}
\FOR |t|=1:|N_time_step|
\text{\color{darkgreen}\,\% leap frog } \\
    |v_phalf| = |v_mhalf| + |a|*|dt|; \\
    |x| \pluseq |v_phalf|*|dt|; \\
    |v| = 0.5*(|v_mhalf|+|v_phalf|);
    |v_mhalf| = |v_phalf|; \\
    \text{\color{darkgreen}\,\% update densities, pressures, accelerations } \\
    |rho| = |GetDensity|( |x|, |m|, |h| ); \\
    |P| = |GetPressure|( |x|, |rho|, |m|, |h| ); \\
    |a| = |GetAcceleration|(|x|, |v|, |m|, |rho|, |P|, |b|, |beta|, |lambda|, |h|);
\END
\end{program}

Our implementation is $\mathcal{O}(N^2)$ because we calculate all pairwise interactions between particles. Alternative implementations have been developed in literature to make the computations more efficient ($\mathcal{O}(N\log N)$.), such as computing interactions only between the $k$ nearest neighbors with using a hierarchical tree algorithm \citep{hernquist1989treesph}.

\subsection{Gross-Pitaevskii-Poisson}\label{sec:poisson}

The resulting equations of motion from self-gravity (Equation~\ref{eqn:poisson}) can be calculated using and $N$-body technique (here $M=1$):
\begin{equation}
\frac{d\mathbf{u}_i}{dt} = \sum_{j\neq i} \frac{m_j(\mathbf{r}_j\mathbf{r}_i)}{(\lvert\mathbf{r}_j\mathbf{r}_i\rvert^2+\epsilon^2)^{3/2}}
\end{equation}
where $\epsilon$ is a smoothing length, used to avoid numerical problems of close particle encounters (where the acceleration blows up) in collision-less dynamics. In SPH, $\epsilon$ is typically equated with the kernel's smoothing length $h$. SPH couples naturally with the Poisson equation, making our method well-suited for finding solutions of the Gross-Pitaevskii-Poisson equations.

\subsection{Adaptive Smoothing Lengths}\label{sec:adaptive}

In some applications it is advantageous to use an adaptive smoothing length $h_i$ for each particle $i$ for improved numerical accuracy. The variable smoothing length depends on the density at the fluid and allows the algorithm to handle regions with high and low densities more precisely \cite{price2007energy}. 

The adaptive smoothing length and the density can be calculated self-consistently using an iterative scheme as follows. The density estimator becomes:
\begin{equation}
\rho_i = \sum_j m_j W(\mathbf{x}_i-\mathbf{x}_j;h_i)
\end{equation}

There is a correction factor to the momentum equation to allow for
the spatial variation in smoothing lengths in the equation of motion (Equation~\ref{eqn:accsph})
\begin{equation}
\frac{d\mathbf{u}_i}{dt} = \sum_j m_j \left( \frac{P_i}{\Omega_i\rho_i^2} + \frac{P_j}{\Omega_j\rho_j^2} \right) \nabla W(\mathbf{x}_i-\mathbf{x}_j;h_i).
\end{equation}
where (for 3D case)
\begin{equation}
\Omega_i = 1+\frac{h_i}{3\rho_i} \sum_j m_j \frac{\partial W(\mathbf{x}_i-\mathbf{x}_j;h_i)}{\partial h_i}.
\end{equation}
An analogous correction correction applies to Equation~\ref{eqn:main}.

The value for $h_i$ is determined by solving for the root of
\begin{equation}
\zeta(h_i) = m_j\left(\frac{\eta}{h_i}\right)^3 - \sum_j m_j W(\mathbf{x}_i-\mathbf{x}_j;h_i)
\end{equation}
using a Newton-Raphson iterator (or alternate technique). $\eta$ is an order unity constant; for our applications, we use $\eta=1.4$. This constraint ensures that $\rho_i h_i^3 = {\rm const}$, i.e., it maintains a constant mass within the smoothing kernel. Given an initial guess for $h_i$, the Newton-Raphson iterator gives an updated value $h_{i,{\rm new}}$ according to:
\begin{equation}
h_{i,{\rm new}} = h_i - \frac{\zeta(h_i)}{\zeta^\prime(h_i)} = h_i  \left( 1 + \frac{\zeta(h_i)}{3\rho_i\Omega_i} \right)
\end{equation}
until a tolerance threshold is reached: $\lvert h_{i,{\rm new}} - h_i \rvert / h_i <\epsilon_{\rm tol}$. We choose $\epsilon_{\rm tol}=10^{-3}$.

\section{Results}\label{sec:results}


Here we present a number of simple numerical tests to demonstrate our SPH method for the Schr\"odinger equation and the NLSE. The aim of these tests is to demonstrate the accuracy of our method (in capturing steady-state solutions and dynamics) and highlight different physical regimes well-suited for our numerical method to handle (such as systems with self-gravity or collapsing solutions).

\begin{figure}[ht]
\begin{center}
\includegraphics[width=3.2in,angle=0,clip=true] {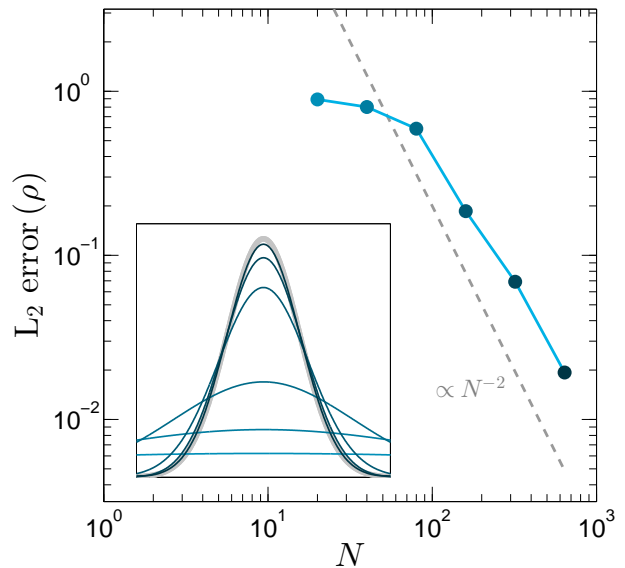}
\caption{Convergence properties of the SPH code for obtaining the ground state of the Schr\"odinger equation with a 1D simple harmonic oscillator potential. Second order convergence is achieved by increasing particle number $N$ (and decreasing smoothing length as $1/N$). The obtained ground state solution for $\rho$ at the various resolutions is shown in the inset (the shades of teal correspond to the shades of the circles in the convergence plot: higher resolution approaches the analytic answer, which is shown with the thick gray line).} 
\label{fig:shoconv}
\end{center}
\end{figure}

\subsection{1D simple harmonic oscillator ground state}\label{ssec:shoconv}

First, we demonstrate the algorithm's ability to recover the ground state of the Schr\"odinger equation with a 1D simple harmonic oscillator potential. This is a very simple test, but can be used to demonstrate our code's convergence properties. The particles are initially drawn from a uniform distribution in the range $[-4,4]$, and we add a damping term with $\gamma=4$ to relax the system to the ground state. We evolve the system in the potential $V(x)=\frac{1}{2}x^2$ until a steady-state configuration is found. We use a smoothing length of $h=200/N$ and a time step of $0.004$. The exact solution is given by
\begin{equation}
\rho_{0} = \pi^{-1/2}{\rm exp} \left(-x^2\right)
\end{equation}

Figure~\ref{fig:shoconv} demonstrates the convergence rate and solution. This method of obtaining the ground state is also useful for generating initial conditions for future tests.

\begin{figure}[ht]
\begin{center}
\includegraphics[width=3.2in,angle=0,clip=true] {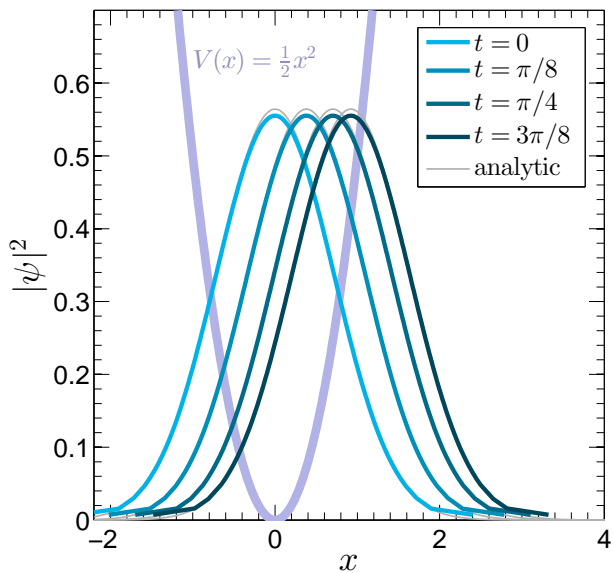}
\caption{(Color online) Time evolution of oscillating wavefunction in simple harmonic oscillator potential. The SPH approach captures the dynamics well. The analytic solution is shown in thin gray lines.} 
\label{fig:shodyn}
\end{center}
\end{figure}

\begin{figure}[ht]
\begin{center}
\includegraphics[width=3.2in,angle=0,clip=true] {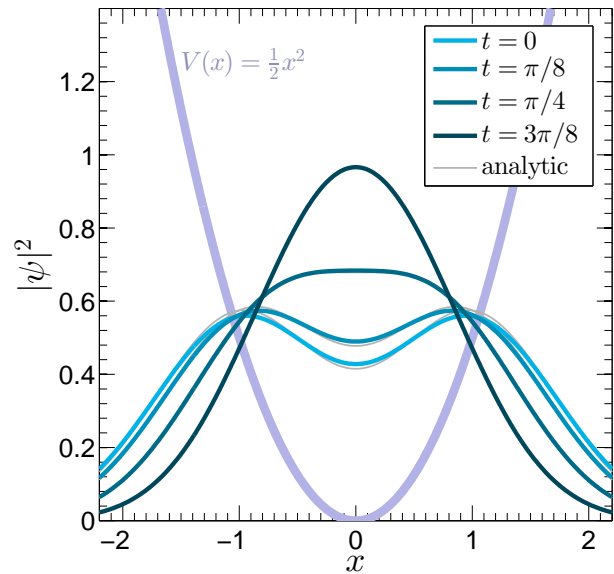}
\caption{(Color online) Evolution of two solitons in simple harmonic oscillator potential. The SPH code captures the profile well, which consists of two solitons oscillating in opposite directions.} 
\label{fig:sscollision}
\end{center}
\end{figure}

\subsection{1D simple harmonic oscillator dynamics}
\label{ssec:shodyn}

We consider the time evolution of a wavefunction in the potential $V(x)=\frac{1}{2}x^2$ with initial conditions:
\begin{equation}
\rho_{0} = \pi^{-1/2}{\rm exp} \left(-x^2\right) ,
\end{equation}
and
\begin{equation}
u_{0} = 1 .
\end{equation}
The evolution has analytic solution
\begin{equation}
\rho(t) = \pi^{-1/2}{\rm exp} \left(-(x-\sin(t))^2\right) .
\end{equation}

Our simulation used $N=300$ particles, smoothing length $h=0.2667$, and time step $dt=0.01$. The initial conditions are obtained by damping a random configuration of particles to the ground state, as in Section~\ref{ssec:shoconv}. Figure~\ref{fig:shodyn} shows the evolution of the wavefunction with time, which matches the analytic solution very well, showing exact agreement with the periodicity of the solution. We note the tiny offset in the peak of the wave function compared to the analytic result is due to the finite number of particles we use, and this discrepancy is reduced with increasing particle number and decreasing smoothing length.

We also demonstrate the dynamics of two solitons in a harmonic potential evolved under the linear Schr\"odinger equation. Two copies of the initial condition in the previous setup are superimposed on top of each other. One soliton is given initial velocity $v=+1$ and the other $v=-1$. In this simple example, the solitons do not interact and just pass through each other. This example demonstrates the algorithm's ability to capture multiple phase dynamics. Our simulation used $N=800$ particles, smoothing length $h=1$, and time step $dt=0.001$. Figure~\ref{fig:sscollision} shows the evolution of the wavefunction with time, which matches the analytic solution very well.

\begin{figure}[ht]
\begin{center}
\includegraphics[width=3.2in,angle=0,clip=true] {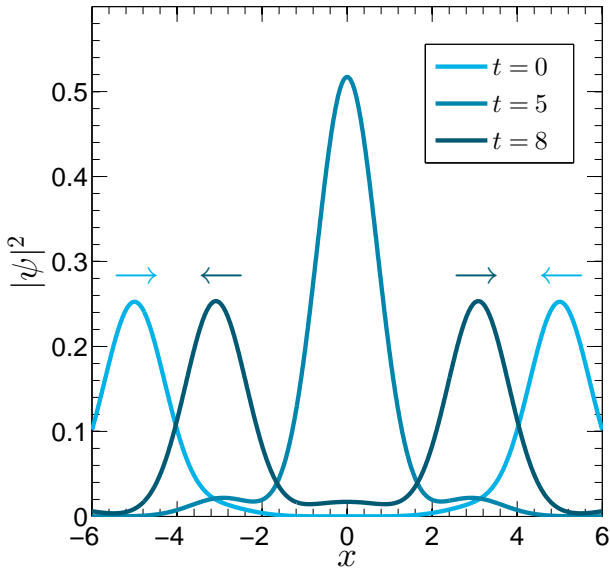}
\caption{(Color online) Collision of two bright solitons evolved under the NLSE. Collision occurs at $t=5$, during which fringes are formed. The solitons return to their original profile after interaction.} 
\label{fig:brightss}
\end{center}
\end{figure}

\begin{figure}[ht]
\begin{center}
\includegraphics[width=3.2in,angle=0,clip=true] {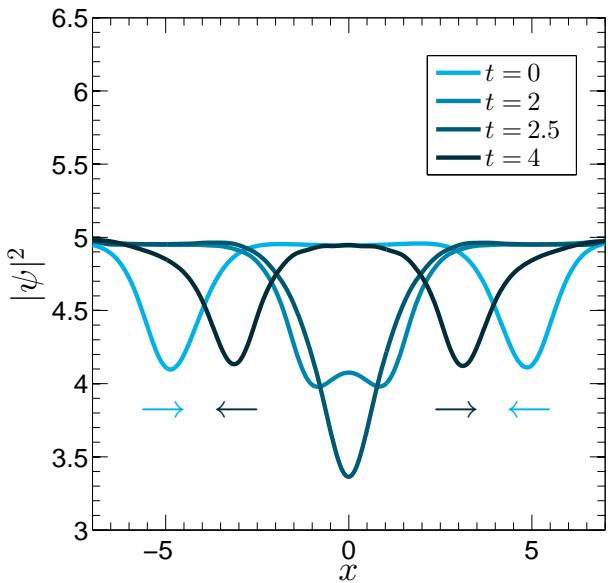}
\caption{(Color online) Collision of two dark solitons evolved under the NLSE. Collision occurs at $t=2.5$.} 
\label{fig:darkss}
\end{center}
\end{figure}

\subsection{1D NLSE soliton-soliton collision}
\label{ssec:sscoll}

We simulate the collision of two solitons that are solution to the NLSE to show the stability and accuracy of the pre- and post- collision nonlinear states with our method. We initialize two bright solitons \cite{bao2013numerical} with analytic profile
\begin{equation}
\psi_0(x) = \frac{1}{2}{\rm sech}(x\pm x_0){\rm e}^{\pm ivx}
\end{equation}
(such an initial condition is initialized by damping an initially uniform distribution of particles under the appropriate potential that gives rise to the profile as the steady-state solution.)
A single bright soliton moves at constant velocity $v$, with peak initially determined by $x=x_0$. We simulate the interaction of two solitons initially located at $x=\pm x_0=\pm 5$ and each traveling with speed $v=1$ towards each other, evolved under the NLSE with $g=-1$. During interaction, oscillations are created, and at collision the peak value of the amplitude doubles from the initial peak values of the single soliton, as expected in soliton-soliton collisions \cite{bao2013numerical}. After interaction, the solitons return to their original profile shape and continue traveling at velocity $v$. The results of our simulation, demonstrating the mentioned behavior, is shown in Figure~\ref{fig:brightss}. The simulations use $N=100$ particles, smoothing length $h=1$, and time step $dt=0.001$. The pre- and post collision profile is preserved very well under evolution: our initial condition has profile peak $0.2529$ and velocity $1$, and the post-collision profile at $t=8$ has peak $0.2526$ and velocity $1.0043$ (note that the analytic solution has initial peak $0.2500$, so there is a small numerical offset in the initial conditions due to the truncation errors associated with relaxing the SPH particles in a potential that yields the initial conditions).

We point out that, as is inherent in SPH, in regions where density is low, the profile may be dominated by the shape of the kernel (e.g. the slight bump in density at $x=0$, $t=8$). Additionally, there are truncation errors in the inital conditions generated by relaxing the SPH particles in a constructed potential to generate the inital conditions. This can potentially lead to additional small amplitude waves propagating in the time evolving solution.

We also demonstrate that the SPH method can capture collision of dark solitons. In the case, the profile we consider is 
\begin{equation}
\psi_0(x) = iv \pm {\rm tanh}(x\pm x_0)
\end{equation}
with initial positions determined by $x=\pm x_0=\pm 5$, and velocity $v=2$. Like the bright soliton, the profile for a single dark soliton remains unchanged under evolution of the NLSE. The velocity of the SPH particles making up the soliton is
\begin{equation}
u(x) = \mp \frac{v {\rm sech}^2(x\pm x_0)}{v^2 +  {\rm tanh}^2(x\pm x_0)}
\end{equation}
Thus, unlike the bright soliton, the initial Lagrangian velocities of the SPH are not constant (note, the soliton profile still moves at constant velocity). Thus this test demonstrates dynamics and ability to maintain a soliton profile under a continuum of phase velocities. The system is evolved by the NLSE with $g=1$. The results of our simulation are shown in Figure~\ref{fig:darkss}. The simulation uses $N=100$ particles, smoothing length $h=0.4$, and time step $dt=0.001$, and periodic boundary conditions on the domain $[-10,10]$. The simulation preserved the profile well after interaction. 

We point out that a general issue with the fluid-like approach of the Madelung equations is that the quantum pressure term can be a problem from the singularity point of view. This is certainly an issue for grid-based methods that attempt to solve the Madelung equations. The density can approach zero and the corresponding velocity can be infinite (so that the momentum is finite). However, with an SPH approach the situation is improved. The velocities are always calculated at the locations of the particles, at which there is always a minimum density determined by the smoothing kernel. Regions with tiny or 0 valued wavefunction have no particles at the location representing the solution.

\subsection{2D BEC ground states}

We calculate the ground state of a 2D BEC by relaxing a random initial condition evolved under the NLSE, with damping damping $\gamma=4$. We calculate the states for BECs with $g=0,10,50,100,250,500$ in a potential $V(x)=\frac{1}{2}\left(x^2+y^2\right)$. The simulations use $N=100$ particles, smoothing length $h=1$, and time step $dt = 0.1$.

\begin{figure}[ht]
\begin{center}
\includegraphics[width=3.2in,angle=0,clip=true] {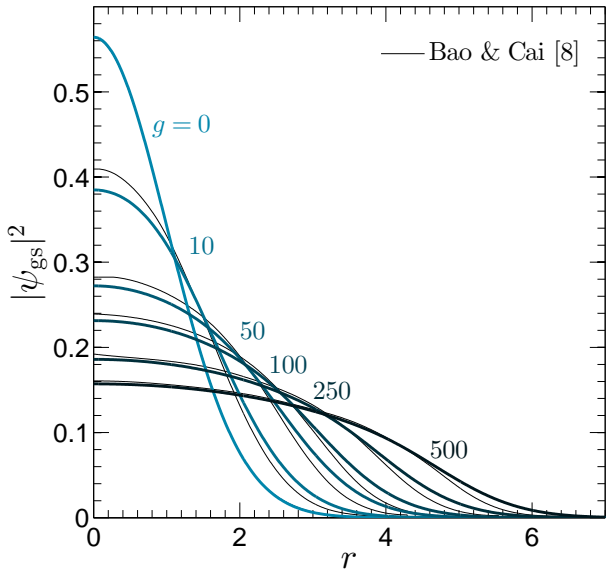}
\caption{(Color online) Ground state of 2D BEC with various values of the parameter $g$, as obtained by the SPH approach and compared to results from high resolution numerical simulations in \protect\cite{bao2013mathematical}.} 
\label{fig:becgs}
\end{center}
\end{figure}

The results are shown in Figure~\ref{fig:becgs}, and are compared to the high resolution numerical simulations in \cite{bao2013mathematical}, showing good agreement.

\begin{figure}[ht]
\begin{center}
\includegraphics[width=3.2in,angle=0,clip=true] {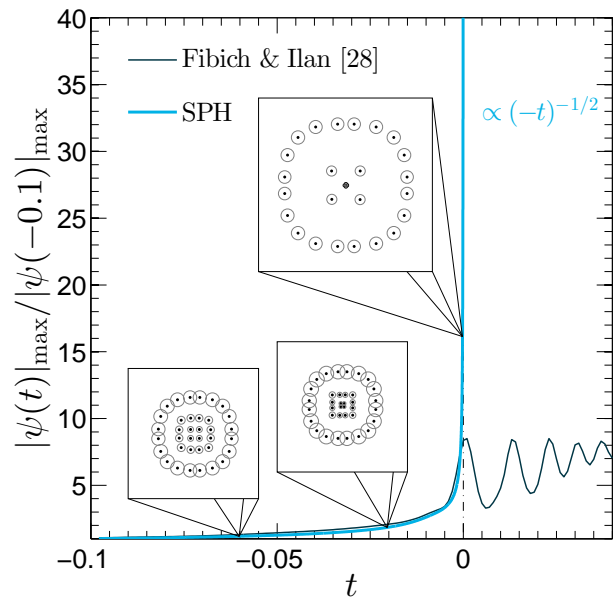}
\caption{(Color online) Comparison of the blowup solution of the focusing NLSE with numerical solutions obtained via our SPH method and a second-order finite difference scheme for comparison. The SPH solution produces a collapse with the right scaling while the finite difference scheme shows focusing-defocusing oscillations early on due to discretization effects. In the insets, we show the location of the SPH particles at three different times, and draw circles around each that have radii proportional to the adaptive smoothing length of the particle. The inset at the third time frame has an aggregation of 12 particles near the singularity.} 
\label{fig:sf}
\end{center}
\end{figure}

\subsection{2D self-focusing NLSE}\label{ssec:sf}

We simulate a blowup solution of the 2D focusing NLSE. The initial condition is given by a Gaussian $\psi=2.99{\rm e}^{-(x^2+y^2)}$. We use $N=36$ particles and adaptive smoothing lengths. Figure~\ref{fig:sf} shows the evolution of the peak of the wavefunction as a function of time, which agrees well with the blowup solution scaling of the problem $\lvert\psi\rvert_{\rm max}\propto (-t)^{-1/2}$ ($t=0$ corresponds to blowup) \cite{fibich2003discretization}. In contrast, a finite-difference approach shows focusing-defocusing oscillations once the focusing becomes too strong \cite{fibich2003discretization} (i.e., the oscillatory behavior of the peak of the wavefunction shown in Figure~\ref{fig:sf}). Additionally, such methods require a large number of resolution elements. Figure~\ref{fig:sf} also shows the configuration of the SPH particles at different times in the simulation, as well as their adaptive smoothing lengths. Many of the particles cluster at the center because of the collapse, which some of the particles are maintained farther out to represent the extent of the wave function at these locations. The inset at showing the particle configuration at the third time frame has an aggregation of $12$ particles near the singularity.

\begin{figure}[ht]
\begin{center}
\includegraphics[width=3.2in,angle=0,clip=true] {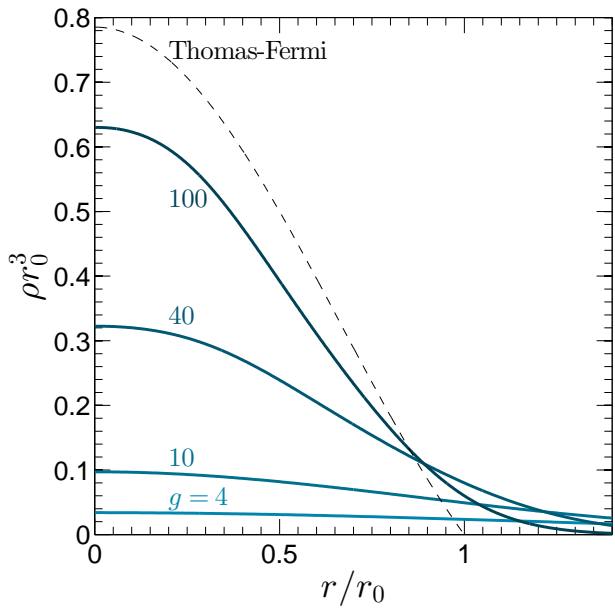}
\caption{(Color online) 3D BEC dark matter halo profiles calculated with our SPH method for various values of the parameter $g$. The profiles approach the Thomas-Fermi approximation for increasing values of $g$.} 
\label{fig:dm}
\end{center}
\end{figure}

\subsection{3D BEC dark matter halo}\label{ssec:dm}

We compute solutions to the Gross-Pitaevskii-Poisson equation that describes self-gravitating BEC dark matter halos by relaxing an initial condition using a damping parameter $\gamma=4$. Our simulations use $N=300$ particles, adaptive smoothing lengths, and units with $G=1$. The solutions, for various values of $g$, are shown in Figure~\ref{fig:dm}. The length scale is normalized by $r_0=\sqrt{\pi c / 4G}$ in the figure. We simulate the cases of $g=4,10,40,100$. In the limit of large $g$, the numerical solution approach the Thomas-Fermi approximation: $\rho\propto {\rm sinc}(\pi r/r_0) $.

The Gross-Pitaevskii-Poisson equation describes one possible physical model for the non-baryonic dark matter that forms a large fraction of the content in our Universe. In this model a fundamental scalar field plays the role of dark matter, and the model is a competitor to the standard $\Lambda$ cold dark matter model \cite{suarez2014review}. Large cosmological simulations with the BEC model for dark matter have been performed on an adaptively-refined mesh to study nonlinear cosmic structure formation of gravitationally collapsed objects \cite{schive2014understanding,schive2014cosmic}. 

The Gross-Pitaevskii-Poisson equation describes other physical systems as well, such as dipolar BECs.

\section{Computational Efficiency}\label{sec:efficiency}

Our SPH approach to find solutions to the NLSE maintains the simplicity and computational efficiency of the original hydrodynamic SPH method. The method only requires that the particle positions, velocities, masses, and smoothing lengths be stored in memory. More advanced techniques, standard in the field of SPH, can be used to make the method $\mathcal{O}(N\log N)$, whereas our simple implementation to calculate pairwise interactions is $\mathcal{O}(N^2)$. The SPH technique has successfully been implemented with over $10^{10}$ particles on standard central processing unit (CPU) clusters with the use of Message Passing Interface (MPI) routines \citep{springel2005cosmological}. In addition, the algorithm is well-suited to the modern graphics processing unit (GPU) and GPU cluster architectures \cite{valdez2012towards,dominguez2013new}, which have shown an order of magnitude increase in efficiency compared to CPU approaches. These methods can simulate a time step of over $10^6$ particles per second. In solving the NLSE, the computations per communication, are increased due to the computation of a pressure tensor rather than a simple pressure, which boosts the computational efficiency of the original hydrodynamic SPH method. A number of numerical methods exist to calculate the self-gravity term, which shows up in the Gross-Pitaevskii-Poisson equation, that are $\mathcal{O}(N\log N)$. These include tree and multipole based methods and will make the subject of future investigations.

\section{Concluding Remarks}\label{sec:conc}

We have demonstrated a new, simple numerical method to solve the NSLE using SPH. The method conserves the normalization condition on the wavefunction to machine-precision. Additionally, the computational domain is unlimited, which is very natural for a wavefunction. The SPH particles that represent the probability density of the wave function automatically adapt to regions where the density is the largest. This makes our method ideal for solving collapsing and singular solutions, an area where standard grid methods face difficulties. One limitation of our method is that the the hydrodynamic equations and the Gaussian kernels are not well-suited for handling systems with singularities, such as scalar quantum vortices (at the vortex core the density is zero), leading to singular hydrodynamic equations \citep{yepez2009vortex}. Investigation of such systems is beyond the scope of the present work, and may require alternate kernel functions to prevent smoothing out the singularities.

The implementation is relatively simple and easily extendable to modifications of the NSLE. The numerical method can be applied to a variety of physical systems, including BECs, nonlinear optics, capillary fluids, dark matter that obeys the Gross-Pitaevskii-Poisson equations, and collapsing singularities.

\section*{Acknowledgments}
This material is based upon work supported by the National Science Foundation Graduate Research Fellowship under grant no. DGE-1144152 (PM). PM would like to thank D. Marsh and X. Guo for careful reading of the manuscript, and H.Y. Schive for valuable discussions.

\bibliography{mybib}{}

\begin{thebibliography}{10}

\bibitem{bradley1995evidence}
C.~C. Bradley, C.~A. Sackett, J.~J. Tollett, and R.~G. Hulet,
\newblock Physical Review Letters {\bf 75}, 1687 (1995).

\bibitem{anderson1995observation}
M.~H. Anderson, J.~R. Ensher, M.~R. Matthews, C.~E. Wieman, and E.~A. Cornell,
\newblock science {\bf 269}, 198 (1995).

\bibitem{bao2014mathematical}
W.~Bao,
\newblock arXiv preprint arXiv:1403.3884  (2014).

\bibitem{anderson1983variational}
D.~Anderson,
\newblock Physical Review A {\bf 27}, 3135 (1983).

\bibitem{gupta1981coupled}
M.~Gupta, B.~Som, and B.~Dasgupta,
\newblock J. Plasma Phys {\bf 25}, 499 (1981).

\bibitem{chavanis2011mass}
P.-H. Chavanis,
\newblock Physical Review D {\bf 84}, 043531 (2011).

\bibitem{bao2003numerical}
W.~Bao, D.~Jaksch, and P.~A. Markowich,
\newblock Journal of Computational Physics {\bf 187}, 318 (2003).

\bibitem{bao2013mathematical}
W.~BAO and Y.~CAI,
\newblock Kinetic \& Related Models {\bf 6} (2013).

\bibitem{antoine2014gpelab}
X.~Antoine and R.~Duboscq,
\newblock gpelab. math. cnrs. fr  (2014).

\bibitem{santos2014comparison}
L.~S. Santos, M.~O. Pires, and D.~Giugno,
\newblock arXiv preprint arXiv:1410.5006  (2014).

\bibitem{Dalfovo96thecondensate}
F.~Dalfovo, L.~Pitaevskii, and S.~Stringari,
\newblock Journal of Research of the National Institute of Standards and
  Technology {\bf 101} (1996).

\bibitem{bao2003ground}
W.~Bao and W.~Tang,
\newblock Journal of Computational Physics {\bf 187}, 230 (2003).

\bibitem{bao2004computing}
W.~Bao and Q.~Du,
\newblock SIAM Journal on Scientific Computing {\bf 25}, 1674 (2004).

\bibitem{bao2014ground}
W.~Bao and Y.~Cai,
\newblock arXiv preprint arXiv:1407.5815  (2014).

\bibitem{succi1996numerical}
S.~Succi,
\newblock Physical Review E {\bf 53}, 1969 (1996).

\bibitem{palpacelli2008quantum}
S.~Palpacelli and S.~Succi,
\newblock Physical Review E {\bf 77}, 066708 (2008).

\bibitem{yepez1998lattice}
J.~Yepez,
\newblock International Journal of Modern Physics C {\bf 9}, 1587 (1998).

\bibitem{yepez2001quantum}
J.~Yepez,
\newblock Physical Review E {\bf 63}, 046702 (2001).

\bibitem{minguzzi2004numerical}
A.~Minguzzi, S.~Succi, F.~Toschi, M.~Tosi, and P.~Vignolo,
\newblock Physics reports {\bf 395}, 223 (2004).

\bibitem{tsekov2009dissipative}
R.~Tsekov,
\newblock arXiv preprint arXiv:0903.0283  (2009).

\bibitem{gingold1977smoothed}
R.~A. Gingold and J.~J. Monaghan,
\newblock Monthly notices of the royal astronomical society {\bf 181}, 375
  (1977).

\bibitem{lucy1977numerical}
L.~B. Lucy,
\newblock The astronomical journal {\bf 82}, 1013 (1977).

\bibitem{monaghan1992smoothed}
J.~J. Monaghan,
\newblock Annual review of astronomy and astrophysics {\bf 30}, 543 (1992).

\bibitem{springel2010smoothed}
V.~Springel,
\newblock Annual Review of Astronomy and Astrophysics {\bf 48}, 391 (2010).

\bibitem{price2012smoothed}
D.~J. Price,
\newblock Journal of Computational Physics {\bf 231}, 759 (2012).

\bibitem{berry1972semiclassical}
M.~V. Berry and K.~Mount,
\newblock Reports on Progress in Physics {\bf 35}, 315 (1972).

\bibitem{fibich2005new}
G.~Fibich, N.~Gavish, and X.-P. Wang,
\newblock Physica D: Nonlinear Phenomena {\bf 211}, 193 (2005).

\bibitem{fibich2003discretization}
G.~Fibich and B.~Ilan,
\newblock Applied numerical mathematics {\bf 44}, 63 (2003).

\bibitem{nelson1994variable}
R.~P. Nelson and J.~C. Papaloizou,
\newblock Monthly Notices of the Royal Astronomical Society {\bf 270}, 1
  (1994).

\bibitem{chaniotis2002remeshed}
A.~Chaniotis, D.~Poulikakos, and P.~Koumoutsakos,
\newblock Journal of Computational Physics {\bf 182}, 67 (2002).

\bibitem{cleary1998modelling}
P.~W. Cleary,
\newblock Applied Mathematical Modelling {\bf 22}, 981 (1998).

\bibitem{fatehi2008discretization}
R.~Fatehi, M.~Fayazbakhsh, and M.~Manzari,
\newblock  {\bf 30}, 243 (2008).

\bibitem{hernquist1989treesph}
L.~Hernquist and N.~Katz,
\newblock The Astrophysical Journal Supplement Series {\bf 70}, 419 (1989).

\bibitem{price2007energy}
D.~Price and J.~Monaghan,
\newblock Monthly Notices of the Royal Astronomical Society {\bf 374}, 1347
  (2007).

\bibitem{bao2013numerical}
W.~Bao, Q.~Tang, and Z.~Xu,
\newblock Journal of Computational Physics {\bf 235}, 423 (2013).

\bibitem{suarez2014review}
A.~Su{\'a}rez, V.~H. Robles, and T.~Matos,
\newblock A review on the scalar field/bose-einstein condensate dark matter
  model,
\newblock in {\em Accelerated Cosmic Expansion}, pp. 107--142, Springer, 2014.

\bibitem{schive2014understanding}
H.-Y. Schive {\em et~al.},
\newblock Physical Review Letters {\bf 113}, 261302 (2014).

\bibitem{schive2014cosmic}
H.-Y. Schive, T.~Chiueh, and T.~Broadhurst,
\newblock Nature Physics {\bf 10}, 496 (2014).

\bibitem{springel2005cosmological}
V.~Springel,
\newblock Monthly Notices of the Royal Astronomical Society {\bf 364}, 1105
  (2005).

\bibitem{valdez2012towards}
D.~Valdez-Balderas, J.~M. Dom{\'\i}nguez, B.~D. Rogers, and A.~J. Crespo,
\newblock arXiv preprint arXiv:1210.1017  (2012).

\bibitem{dominguez2013new}
J.~M. Dom{\'\i}nguez, A.~J. Crespo, D.~Valdez-Balderas, B.~D. Rogers, and
  M.~G{\'o}mez-Gesteira,
\newblock Computer Physics Communications {\bf 184}, 1848 (2013).

\bibitem{yepez2009vortex}
J.~Yepez, G.~Vahala, and L.~Vahala,
\newblock The European Physical Journal-Special Topics {\bf 171}, 9 (2009).

\end{thebibliography}

\vfill\eject

\end{document}